\documentclass[english,pra, twocolumn, showpacs, showkeyw]{revtex4-1}
\usepackage[T1]{fontenc}
\usepackage[latin9]{inputenc}
\usepackage{textcomp}
\usepackage{amsmath}
\usepackage{amssymb}
\usepackage{stackrel}
\usepackage{graphicx}

\makeatletter

\newcommand{\lyxdot}{.}

\usepackage[colorlinks = true,
            linkcolor =red,
            urlcolor  =magenta,
            citecolor = blue,
            anchorcolor = blue]{hyperref}

\makeatother

\usepackage{babel}
\begin{document}
\title{Separating the Wave and Particle Attributes of Two Entangled Photons }
\author{Yusuf Turek}
\email{yusuftu1984@hotmail.com}

\author{Yi-Fang Ren }
\affiliation{School of Physics, Liaoning University, Shenyang, Liaoning 110036,
China}
\date{\today}
\begin{abstract}
Wave-particle duality is one of the most intriguing counterfactual
concepts in quantum theory. In our common sense, the wave and particle
properties of a quantum object are inseparable. However, the recent
studies based on Quantum Cheshire Cat phenomena showed that separating
the physical properties of a quantum object including wave and particle
attributes from itself are possible in microscopic system described
by two-state vector formalism. In this study, we put forward a feasible
scheme to spatially separate the wave and particle attributes of two
entangled photons by properly choosing the pre- and post-selection
of path states. Our scheme also guarantees that the observation of
wave and particle properties of the two entangled photons always obey
the Bohr's complementarity principle.
\end{abstract}
\maketitle

\section{\label{sec:1}Introduction}

Measurement problems plays essential roles for our understanding of
the nature of the quantum theory and leads to many counterintuitive
quantum phenomena. One of them is the wave-particle duality of quantum
objects which depends on measurement apparatus \citep{1984}. In turn,
quantum objects display its wave- or particle-like behavior depending
on the experimental technique used to measure them \citep{1986,1997K,1998,PhysRevLett.91.130401,2019}.
The fundamental issues related to wave-particle duality and Bohr's
complementarity principle \citep{1928,1979,1991S} can be exemplified
by Young-like double-slit experiments and their variants, e.g. Wheeler's
delayed choice experiment \citep{1978}. It has been implemented in
variety of systems such as photons, atoms, and superconducting circuits
\citep{PhysRevLett.77.2154,2007,2008,RN2185,2015M,2012G,2012A,2017K,RevModPhys.88.015005,PhysRevA.35.2532,2020D}.
Furthermore, the particle and wave-like behaviors of a single photon
\citep{2011} and multi-entangled photons \citep{2019} has also been
observed simultaneously in one and the same experimental setup \citep{2012Ad,2014Sh}.
Those innovative works provides a new perspective for the in-depth
understanding of wave-particle duality.

In general, since the wave and particle properties of a quantum object
coexist simultaneously, one may always suppose to that the two attributes
of single entity cannot be separated from each other. However, recent
research results on the topic \citep{2021W,2023} has shown, upon
exploiting the idea of Quantum Cheshire Cat (QCC) \citep{Aharonov_2013},
that one may spatially separates of wave and particle attributes of
a quantum object. The QCC is a fictitious character in the famous
novel ``Alice in Wonderland''. As described in that novel, the Cheshire
cat has the ability of disappearing in space while its grin is still
visible. However, in our usual life, we can imagine a cat without
a grin, but it beyond our common sense to accept the existence of
a grin without a cat as Alice said : ``Well! I've often seen a cat
without a grin, but a grin without a cat! It's the most curious thing
I ever saw in my life!''.

The QCC proposal introduced by Y. Aharonov and his collaborators \citep{Aharonov_2013}
is aim to detach a quantum object from its inherent features. QCC
cat may exist in a quantum system described by two-state vector formalism
\citep{PhysRev.134.B1410,PhysRevA.41.11,Aharonov2008}. Furthermore,
its realization depends on performing a post-selected weak measurement
\citep{59,70,RN2035,RN2066,71} to extract the weak values \citep{PhysRevA.72.052111}
of the observables associated to the quantum object under investigation.

After the original QCC scheme proposed, it have received large attention
and has been investigated theoretically \citep{guryanova2012complete,Matzkin_2013,dilorenzo2013hunting,RN2193,ibnouhsein2014twin,Corr2015,DUPREY20181,das2019teleporting,2020A,PhysRevA.103.012228,RN2192,2021W,Hance_2023,hance2023dynamical,zhou2023quantum,PhysRevA.107.052214}
and experimentally \citep{Atherton:15,PhysRevA.94.012102,RN2190,2021Y,RN2191,danner2023threepath,2023,RN2196}.
Among the different studies, a protocol to separate the wave and particle
attributes of a single photon has been proposed \citep{2021W} and
implemented \citep{2023}. Furthermore, in \citep{2017}, it has been
experimentally observed that two photons can be cast in a wave-particle
entangled state provided that suitable initial entangled polarization
states are injected into the apparatus. Although the separation of
particle and wave-like behaviors of a single photon has been observed
theoretically and experimentally, the possibility of such observation
regarding multi-photon states remains to be explored. Since the entanglement
is a key resource for quantum information and quantum communication,
a question arises on whether the separation of wave- particle attributes
of two entangled photons is possible. As we will see in the following,
our results provide an affirmative answer to this question.

In this paper, motivated by previous studies \citep{2017,2021W,2023},
we introduce a feasible theoretical model to spatially separate wave
and particle attributes of two entangled photons based on QCC phenomena.
After appropriate pre- and post-selection of the paths of photons,
we can find the wave and particle features of two entangled photons
in different places. Besides their fundamental interest in the understanding
of counterfactual communication, our results may promote the development
of quantum metrology.

The remaining part of this paper is organized as follows. In Section
\ref{sec:2},we briefly introduce the idea of quantum Cheshire cat.
In Section\ref{sec:3}, we describe in details our proposal and show
that by suitable pre- and post-selection one can realize the separation
of wave and particle attributes of two entangled photons. In the same
Section, we also describe a possible implementation of our scheme
on an optical platform. Finally, in Section \ref{sec:4}, we close
the paper with some concluding remarks and discussions.

\section{\label{sec:2} Brief introduction of QCC }

In general, an object and its inherent properties always coexist and
cannot be separated at anytime. In original proposal of Aharonov and
collaborators \citep{Aharonov_2013}, they found that the polarization
(say, the grin) property can be separated from a photon (say, the
cat) itself in a pre- and post-selected system. Here, we briefly summarize
the ingredients required to observe such phenomena. Initially, a horizontally
polarized photon enters a Mach-Zehnder interferometer (MZI) and passes
through its right and left arms with equal probability, and the corresponding
state reads as
\begin{equation}
\vert\phi_{i}\rangle=\frac{1}{\sqrt{2}}\left(i\vert L\rangle+\vert R\rangle\right)\vert H\rangle,\label{eq:1}
\end{equation}
where $\vert L\rangle$ and $\vert R\rangle$ denote the left and
right arms of MZI, i.e. the two possible paths of the photon, and
$\vert H\rangle$ denotes the horizontal polarization of the photon.
We choose the state $\vert\phi_{i}\rangle$ as our preselected state.
By adding and adjusting some optical elements, we may post-select
the paths and polarization of the photon to obtain the following state
vector
\begin{equation}
\vert\phi_{f}\rangle=\frac{1}{\sqrt{2}}\left(\vert L\rangle\vert H\rangle+\vert R\rangle\vert V\rangle\right),\label{eq:2}
\end{equation}
where $\vert V\rangle$ represents the vertical polarization of the
photon. Now, we measure the position (path) of the photon together
with its circular polarization inside the MZI. In the two-state-vector
formalism, the weak value of an observable $C$ is given by
\begin{equation}
\langle C\rangle_{w}=\frac{\langle\phi_{f}\vert C\vert\phi_{i}\rangle}{\;\langle\phi_{f}\vert\phi_{i}\rangle}.\label{eq:3}
\end{equation}
This weak value can describe the true value of observable $C$ in
time interval corresponded to pre-and postselected states $\vert\phi_{i}\rangle$
and $\vert\phi_{f}\rangle$. If we want to determine the location
of the photon and its polarization in left or right arms of MZI in
the region from $\vert\phi_{i}\rangle$ to $\vert\phi_{f}\rangle$,
we should define the corresponding observables to those quantities.
Explicitly, we have the operators
\begin{equation}
\Pi_{\mu}=\vert\mu\rangle\langle\mu\vert,\label{eq:4}
\end{equation}

\begin{equation}
\sigma_{z}^{\mu}=\sigma_{z}\otimes\Pi_{\mu}.\label{eq:5}
\end{equation}
Here, $\Pi_{L}\left(\Pi_{R}\right)$ with $\mu\in\left(L,R\right)$
is the projection operator of left (right) arm of MZI and $\sigma_{z}=\vert\uparrow_{y}\rangle\langle\uparrow_{y}\vert-\vert\downarrow_{y}\rangle\langle\downarrow_{y}\vert$
is circular polarization observable of the photon with $\vert\uparrow_{y}\rangle=\frac{1}{\sqrt{2}}\left(\vert H\rangle+i\vert V\rangle\right)$
and $\vert\downarrow_{y}\rangle=\frac{1}{\sqrt{2}}\left(\vert H\rangle-i\vert V\rangle\right)$.
The weak values of $\Pi_{\mu}$ and $\sigma_{z}^{\mu}$ can characterize
the definite values of exact location of spatial degree of freedom
and circular polarization of the photon inside the MZI, respectively.

By taking into account the above pre- and post-selected states $\vert\phi_{i}\rangle$
and $\vert\phi_{f}\rangle$, the weak value of $\Pi_{\mu}$ and $\sigma_{z}^{\mu}$
are given by
\begin{equation}
\langle\Pi_{\mu}\rangle_{w}=\delta_{\mu L},\label{eq:6}
\end{equation}
and
\begin{equation}
\langle\sigma_{z}^{\mu}\rangle_{w}=\delta_{\mu R},\label{eq:7}
\end{equation}
respectively. Here, $\delta_{ij}$ denote the Kronecker $\delta$-function
which defined as
\begin{equation}
\delta_{ij}=\begin{cases}
1, & i=j\\
0. & i\neq j
\end{cases}\label{eq:8}
\end{equation}
 We can see that $\langle\Pi_{L}\rangle_{w}=\langle\sigma_{z}^{R}\rangle=1$
and other weak values all equal to zero. The weak values in Eq.(\ref{eq:6})
and Eq. (\ref{eq:7}) indicated that the spatial degree freedom of
the photon is located in the left arm (photon itself go through the
left arm of MZI), while its circular polarization is detected with
unit probability in the right arm of the MZI. This means that in properly
pre- and post-selected quantum systems one can realize the separation
of the object itself from its inherent properties, as the Cheshire
cat.

\section{\label{sec:3}Separation of wave-particle properties of two entangled
photons }

\begin{figure}
\includegraphics[scale=0.6]{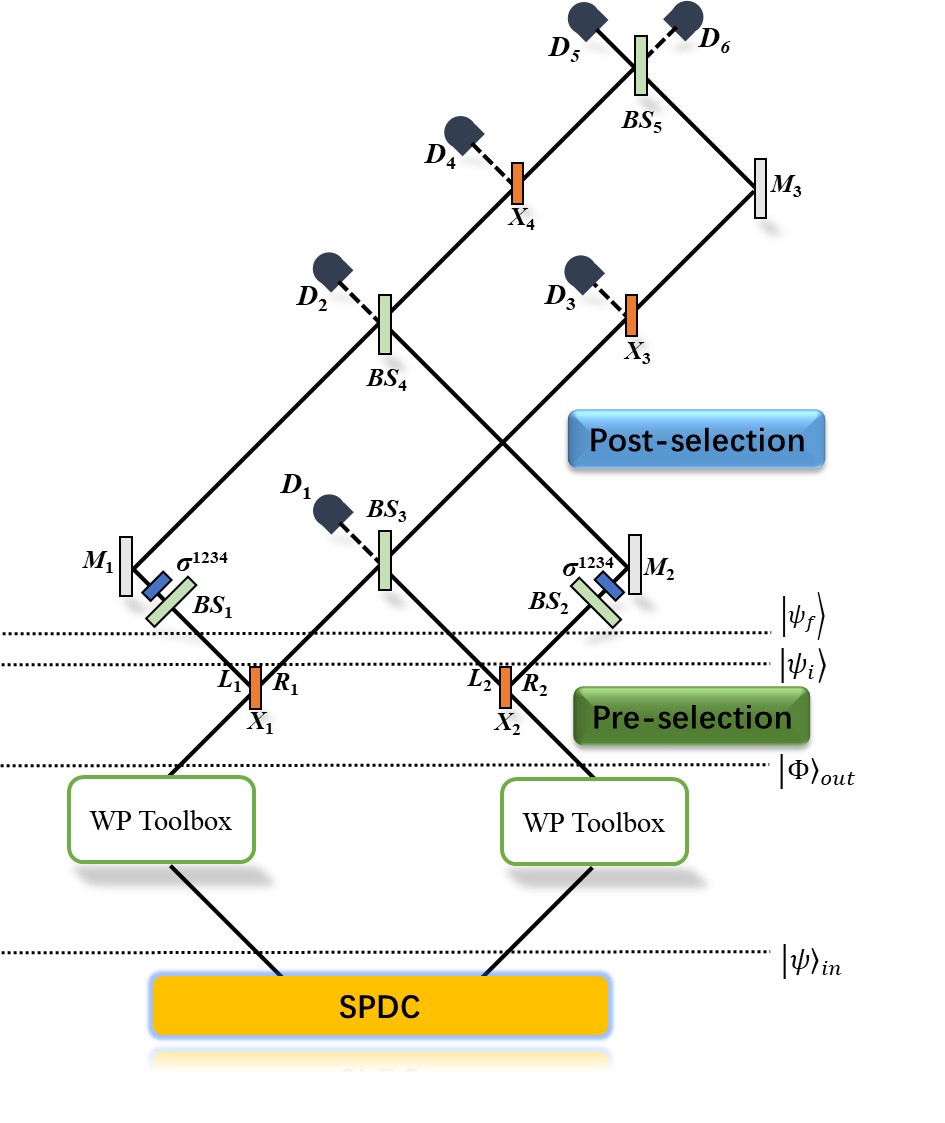}

\caption{\label{fig:1} Illustration of separating the wave and particle attributes
 of two entangled photons.}
\end{figure}

In this section, we describe our proposal for the separation of wave-particle
properties of two entangled photons. The schematic diagram of our
proposed setup is shown in Fig. \ref{fig:1}. The one of most important
points of our scheme is the preparation of so-called wave-particle
(WP) toolbox introduced in Ref. \citep{2017} to convert superpositions
of polarization states to the superpositions of path and polarization
states. Next we describe our scheme step by step along with our schematic
diagram depicted in Fig. \ref{fig:1}.

We assume that in type-I spontaneous parametric down-conversion (SPDC)
process two photons with same polarization have generated, i.e. $\vert V\rangle_{1}\vert V\rangle_{2}$
or $\text{\ensuremath{\vert H\rangle_{1}\vert H\rangle_{2}}}$. Upon
using a half-wave plate (HWP, not shown in the figure) rotated at
an angle $\alpha/2$ on the path of both the photons, the two-photon
state becomes

\begin{equation}
\vert\psi\rangle_{in}=\cos\alpha\vert V\rangle\vert V\rangle+\sin\alpha\vert H\rangle\vert H\rangle,\label{eq:22}
\end{equation}
where $\alpha$ is an adjustable parameter. After this stage, each
photon enters one of two identical WP toolboxes, and the final output
state can be written as \citep{2017}
\begin{align}
\vert\Phi\rangle_{out} & =\cos\alpha\vert W\rangle_{1}\vert W^{\prime}\rangle_{2}+\sin\alpha\vert P\rangle_{1}\vert P^{\prime}\rangle_{2}\nonumber \\
 & =\cos\alpha\vert W\rangle\vert W^{\prime}\rangle+\sin\alpha\vert P\rangle\vert P^{\prime}\rangle,\label{eq:9-1}
\end{align}
where \begin{subequations}
\begin{align}
\vert W\rangle & \equiv\vert Wave\rangle=e^{i\frac{\phi_{1}}{2}}\left(\cos\frac{\phi_{1}}{2}\vert1\rangle-i\sin\frac{\phi_{1}}{2}\vert3\rangle\right),\label{eq:10-1-1}\\
\vert W^{\prime}\rangle & \equiv\!\!\vert Wave^{\prime}\rangle=e^{i\frac{\phi_{1}}{2}}\left(\cos\frac{\phi_{1}}{2}\vert1^{\prime}\rangle\!-i\sin\frac{\phi_{1}}{2}\vert3^{\prime}\rangle\right),\label{eq:11-1-1}\\
\vert P\rangle & \equiv\vert Particle\rangle=\frac{1}{\sqrt{2}}\left(\vert2\rangle+e^{i\phi_{1}}\vert4\rangle\right),\label{eq:12-1-1}\\
\vert P^{\prime}\rangle & \equiv\vert Particle\rangle=\frac{1}{\sqrt{2}}\left(\vert2^{\prime}\rangle+e^{i\phi_{1}}\vert4^{\prime}\rangle\right),\label{eq:13-1-1}
\end{align}
 \end{subequations}with $\vert n\rangle$ and $\vert n^{\prime}\rangle$
( $n,n^{\prime}\in\{1,2,3,4\}$ ) being the $n(n^{\prime})$-th output
modes of from the two WP toolboxes \citep{2017}, and $\phi_{1}$,
$\phi_{1}^{\prime}$, $\phi_{2}$ and $\phi_{2}^{\prime}$ being controllable
phase shifts in the two WP toolboxes. Notice that these states can
characterize the capabilities ($\vert W\rangle$and $\vert W^{\prime}\rangle$)
and incapabilities ($\vert P\rangle$ and $\vert P^{\prime}\rangle$)
of the two photons to produce interference. In \citep{2017}, the
generation processes of superpositions of wave and particle states
for single photon and two entangled photons both have been analyzed.
For more details about the generation of the state, i.g, Eqs. (\ref{eq:10-1-1}-\ref{eq:13-1-1})
in the lab we refer to the original paper \citep{2017}.

Finally, as shown in Fig. \ref{fig:1}, one detects the two photons
simultaneously by measuring the "wave operator" characterized by
the projectors $X_{1}=\vert W\rangle\langle W\vert$ and $X_{2}=\vert W^{\prime}\rangle\langle W^{\prime}\vert$.
The action of those operators is such that only the wave states $\vert W\rangle$
and $\vert W^{\prime}\rangle$ are transmitted whereas the particle
states $\vert P\rangle$ and $\vert P^{\prime}\rangle$ are reflected.
In this way, the state $\vert\Phi\rangle_{out}$ is transformed into

\begin{align}
\vert\psi_{i}\rangle & =\cos\alpha\vert R_{1}L_{2}\rangle\vert WW^{\prime}\rangle+\sin\alpha\vert L_{1}R_{2}\rangle\vert PP^{\prime}\rangle\nonumber \\
 & =\cos\alpha\vert R_{1}L_{2}WW^{\prime}\rangle+\sin\alpha\vert L_{1}R_{2}PP^{\prime}\rangle.\label{eq:W}
\end{align}
This is our pre-selected state. Here, as the notation of QCC, we use
the composite system of path and attributes of two photons as $\vert.\ .\rangle_{paths}\vert.\ .\rangle_{attributes}$.
$\vert L_{i}\rangle$ and $\vert R_{i}\rangle$ denoting the left
and right paths, and $i=1,2$ labels the first and second photons.
$\vert R_{1}L_{2}\rangle\vert WW^{\prime}\rangle$ represents $1$st
(2nd) photon's wave property $\vert W\rangle$ ($\vert W^{\prime}\rangle$)
at $\vert R_{1}\rangle$ ($\vert L_{2}\rangle$), and $\vert L_{1}R_{2}\rangle\vert PP{}^{\prime}\rangle$
represents $1$st (2nd) photon's particle property $\vert P\rangle$
($\vert P^{\prime}\rangle$) at $\vert L_{1}\rangle$ ($\vert R_{2}\rangle$).

In order to realize the separation of wave and particle attributes
of two entangled photons, the post-selected state should be of the
form
\begin{equation}
\vert\psi_{f}\rangle=\frac{1}{\sqrt{2}}\left(\vert R_{1}L_{2}WW^{\prime}\rangle+\vert L_{1}R_{2}PP^{\prime}\rangle\right).\label{eq:14-1}
\end{equation}
This means that in our scheme the post-selected state is a special
case of the pre-selected class of states with $\alpha=\frac{\pi}{4}$.

The determination of post-selected state $\vert\psi_{f}\rangle$ is
crucial for the implementation of our scheme. As shown in Fig. \ref{fig:1},
its verification can accomplished by an optical setup made of five
beam splitters $BS_{i}$ ($i=1,2,3,4,5$), two $\sigma^{1234}$ operators,
two $X_{i}$ operators, three mirrors $M_{i}$ ($i=1,2,3$), and six
detectors $D_{i}$ ($i=1,2,3,4,5,6$). The successful post-selection
occurs iff the detector $D_{5}$ click. On the contrary, if our post-seleceted
state to be in $\vert\psi_{f}\rangle$, then, in the optical arrangement
of Fig. \ref{fig:1}, we can guarantee only the detector $D_{5}$
give the answer ,``yes''. Let us prove this statement.

At first, the photons in $L_{1}$ and $R_{2}$ arms simultaneously
go through the beam splitters $BS_{1}$ and $BS_{2}$ followed by
$\sigma^{1234}$ operators. Here, the effect of $\sigma^{1234}$ operator
is to switch the photon paths in the four-dimensional basis $\{\vert1\rangle,\vert2\rangle,\vert3\rangle,\vert4\rangle\}$.
Its matrix form can be written as \citep{2017}
\begin{equation}
\sigma^{1234}=\left(\begin{array}{cccc}
0 & \ 1\  & 0\  & 0\\
1 & 0 & 0 & 0\\
0 & 0 & 0 & 1\\
0 & 0 & 1 & 0
\end{array}\right),\label{eq:22-1}
\end{equation}
and its base vectors can be written as \begin{subequations}
\begin{align}
\vert1\rangle & =(\begin{array}{cccc}
1 & 0 & 0 & 0\end{array})^{T},\vert2\rangle=(\begin{array}{cccc}
0 & 1 & 0 & 0\end{array})^{T},\\
\vert3\rangle & =(\begin{array}{cccc}
0 & 0 & 1 & 0\end{array})^{T},\vert4\rangle=(\begin{array}{cccc}
0 & 0 & 0 & 1\end{array})^{T}.
\end{align}

\end{subequations} The action of $\sigma^{1234}$ is to switch paths
as follows \citep{2023}
\begin{equation}
\vert1\rangle\stackrel[\sigma^{1234}]{\sigma^{1234}}{\rightleftharpoons}\vert2\rangle,\ \vert3\rangle\stackrel[\sigma^{1234}]{\sigma^{1234}}{\rightleftharpoons}\vert4\rangle.
\end{equation}
The beam splitters $BS_{1}$ and $BS_{2}$ are adjusted to perform
the transformations as \begin{subequations}
\begin{align}
\vert2\rangle & \overset{BS_{1}}{\longrightarrow}\frac{1}{\sqrt{2}}(\vert2\rangle+\vert4\rangle)\\
\vert4\rangle & \overset{BS_{1}}{\longrightarrow}\frac{1}{\sqrt{2}}(\vert2\rangle-\vert4\rangle)
\end{align}
 and
\begin{align}
\vert2^{\prime}\rangle & \overset{BS_{1}}{\longrightarrow}\frac{1}{\sqrt{2}}(\vert2^{\prime}\rangle+\vert4^{\prime}\rangle)\\
\vert4^{\prime}\rangle & \overset{BS_{1}}{\longrightarrow}\frac{1}{\sqrt{2}}(\vert2^{\prime}\rangle-\vert4^{\prime}\rangle)
\end{align}
 \end{subequations} In this way, the particle state can converted
to the wave state after the action of $BS_{1}$ ($BS_{2}$) and $\sigma^{1234}$,
i.e,
\begin{align}
\vert P\rangle\overset{BS_{1}}{\longrightarrow} & \frac{1}{\sqrt{2}}\left[\frac{1}{\sqrt{2}}(\vert2\rangle+\vert4\rangle)+e^{i\phi_{1}}\frac{1}{\sqrt{2}}(\vert2\rangle-\vert4\rangle)\right]\nonumber \\
= & e^{i\frac{\phi_{1}}{2}}\left(\cos\frac{\phi_{1}}{2}\vert2\rangle-i\sin\frac{\phi_{1}}{2}\vert4\rangle\right)\overset{\sigma^{1234}}{\longrightarrow}\nonumber \\
 & e^{i\frac{\phi_{1}}{2}}\left(\cos\frac{\phi_{1}}{2}\vert1\rangle-i\sin\frac{\phi_{1}}{2}\vert3\rangle\right)=\vert W\rangle.\label{eq:25}
\end{align}
The same happens for primed paths with $\vert n^{\prime}\rangle\in\left(\vert1^{\prime}\rangle,\vert2^{\prime}\rangle,\vert3^{\prime}\rangle,\vert4^{\prime}\rangle\right)$,
as
\begin{equation}
\vert P^{\prime}\rangle\overset{BS_{1}}{\longrightarrow}e^{i\frac{\phi_{1}^{\prime}}{2}}\left(\cos\frac{\phi_{1}^{\prime}}{2}\vert2^{\prime}\rangle-i\sin\frac{\phi_{1}^{\prime}}{2}\vert4^{\prime}\rangle\right)\overset{\sigma^{1234}}{\longrightarrow}\vert W^{\prime}\rangle\label{eq:26}
\end{equation}
 Thus, after the photons in $L_{1}$ and $R_{2}$ arms simultaneously
go through beam splitters $BS_{1}$ and $BS_{2}$ followed by the
$\sigma^{1234}$ operators, our post-selected state $\vert\psi_{f}\rangle$
changed to
\begin{equation}
\vert\psi_{f^{\prime}}\rangle=\frac{1}{\sqrt{2}}\left(\vert R_{1}L_{2}\rangle+\vert L_{1}R_{2}\rangle\right)\vert WW^{\prime}\rangle.\label{eq:27}
\end{equation}
 Next, by adjusting the beam-splitters $BS_{3}$ and $BS_{4}$, respectively,
such that states $\vert R_{1}\rangle$ and $\vert L_{1}\rangle$ ($\vert L_{2}\rangle$
and $\vert R_{2}\rangle$) totally transmitted (reflected) to guarantee
$\vert R_{1}L_{2}\rangle$ and $\vert L_{1}R_{2}\rangle$ go through
$X_{3}$ and $X_{4}$. Under this action the detectors $D_{1}$ and
$D_{2}$ both will not click. In this step, the modified form of $\vert\psi_{f^{\prime}}\rangle$
can be written as
\begin{equation}
\vert\psi_{f^{\prime}}\rangle\overset{BS_{3},BS_{4}}{\longrightarrow}\vert\psi_{f^{\prime\prime}}\rangle=\frac{1}{\sqrt{2}}\left(\vert R\rangle+\vert L\rangle\right)\vert\mathcal{\boldsymbol{W}}\rangle.\label{eq:28}
\end{equation}
Here, we denote $\vert R_{1}L_{2}\rangle\overset{BS_{3}}{\longrightarrow}\vert R\rangle$,
$\vert L_{1}R_{2}\rangle\overset{BS_{4}}{\longrightarrow}\vert L\rangle$
and $\vert\mathcal{\boldsymbol{W}}\rangle=\vert WW^{\prime}\rangle$.
The $X_{3}=X_{4}=\vert W\rangle\langle W\vert$ operators transmit
only the wave state, and this action guarantees that detectors $D_{3}$
and $D_{4}$ will not click. The beam splitter $BS_{5}$ is chosen
as $\vert R\rangle\overset{BS_{5}}{\longrightarrow}\frac{1}{\sqrt{2}}\left(\vert R\rangle+\vert L\rangle\right)$,
$\vert L\rangle\overset{BS_{5}}{\longrightarrow}\frac{1}{\sqrt{2}}\left(\vert R\rangle-\vert L\rangle\right)$,
such that
\begin{equation}
\vert\psi_{f^{\prime\prime}}\rangle\overset{BS_{5}}{\longrightarrow}\vert\Psi_{f}\rangle=\vert R\rangle\vert\mathcal{\boldsymbol{W}}\rangle.\label{eq:29}
\end{equation}
 Hence, we can confirm that detector $D_{5}$ is click with certainty.
Contrarily, if and only if the detector $D_{5}$ click with $100\%$
probability, then our schematics also can prove that the postslected
state should be chosen to $\vert\psi_{f}\rangle$.

After choosing the pre- and post-selected state of our scheme, we
have to extract the weak values to achieve our goal. The operators
that measure whether the wave and particle attributes of two entangled
photons are present in the left and right arms is defined as
\begin{equation}
\Pi_{\nu}^{\mu_{i}}=\vert\mu_{i}\rangle\langle\mu_{i}\vert\otimes\vert\nu\rangle\langle\nu\vert,
\end{equation}
with $\nu\in\{W,W^{\prime},P,P^{\prime}\}$, $\mu_{i}\in\{L_{i},R_{i}\}$
with $i=1,2$. Above every operators has its meaning, such as $\Pi_{W}^{L_{1}}=\vert L_{1}\rangle\langle L_{1}\vert\otimes\vert W\rangle\langle W\vert$
can describe the findings of the wave attribute of first photon on
the left arm and $\Pi_{P^{\prime}}^{R_{2}}=\vert R_{2}\rangle\langle R_{2}\vert\otimes\vert P^{\prime}\rangle\langle P^{\prime}\vert$
can describe the findings of the particle attribute of second photon
on the right arm, etc.

By substituting the above-defined pre- and post-selected states $\vert\psi_{f}\rangle$
and $\vert\psi_{f}\rangle$ into Eq. (\ref{eq:3}), we obtain the
corresponding weak values of the observables $\Pi_{\nu}^{\mu_{i}}$,
i.e., \begin{subequations}
\begin{align}
\langle\Pi_{W}^{\mu_{i}}\rangle_{w} & =\frac{\cos\alpha}{\cos\alpha+\sin\alpha}\delta_{1i}\delta_{R_{i}\mu_{i}},\\
\langle\Pi_{P}^{\mu_{i}}\rangle_{w} & =\frac{\sin\alpha}{\cos\alpha+\sin\alpha}\delta_{1i}\delta_{L_{i}\mu_{i}},\\
\langle\Pi_{W^{\prime}}^{\mu_{i}}\rangle_{w} & =\frac{\cos\alpha}{\cos\alpha+\sin\alpha}\delta_{2i}\delta_{L_{i}\mu_{i}},\\
\langle\Pi_{P^{\prime}}^{\mu_{i}}\rangle_{w} & =\frac{\sin\alpha}{\cos\alpha+\sin\alpha}\delta_{2i}\delta_{R_{i}\mu_{i}}.
\end{align}
 \end{subequations} This is the main result of our study, and they
can characterize the exact values of operators $\Pi_{\nu}^{\mu_{i}}$
in the region from $\vert\psi_{i}\rangle$ to $\vert\psi_{f}\rangle$.
If $\alpha\neq0,\frac{\pi}{2}$, these weak values indicated that
the wave and particle attributes of two entangled photons are spatially
separated, and in each one of the four arms we only have one attribute
(wave or particle ) of the two photons. In the particular case, $\alpha=\frac{\pi}{4},$
the weak values for the wave and particle attributes may be written
as $\langle\Pi_{P}^{L_{1}}\rangle_{w}=\langle\Pi_{W}^{R_{1}}\rangle_{w}=\frac{1}{2}$,
$\langle\Pi_{P^{\prime}}^{R_{2}}\rangle_{w}=\langle\Pi_{W^{\prime}}^{L_{2}}\rangle_{w}=\frac{1}{2}$.
This means that for $\alpha=\frac{\pi}{4}$ case, half of the particle
attributes of the two entangled photons occurred in $L_{1}$ and $R_{2}$
arms and half of the wave attributes are present in $R_{1}$ and $L_{2}$
arms of the interferometer, respectively. Furthermore, it is clear
that these weak values cannot take arbitrary large values, and always
obey the Bohr's complementarity principle, i.e.,
\begin{align}
\langle\Pi_{P}^{L_{1}}\rangle_{w}+\langle\Pi_{W}^{R_{1}}\rangle_{w} & =1,\\
\langle\Pi_{P^{\prime}}^{R_{2}}\rangle_{w}+\langle\Pi_{W^{\prime}}^{L_{2}}\rangle_{w} & =1.
\end{align}
 In this proposal, the correctly choosing of post-selected state is
very crucial and we only focus on the cases of $D_{5}$ click with
certainty. Otherwise, if any other detector clicks, the post-selection
fails, and we fail to separate the wave and particle attributes of
two entangled photons.

\section{\label{sec:4} CONCLUSION AND DISCUSSION}

In this paper, we have investigated the possibility of spatially separating
the wave and particle features of two entangled photons. Our work
is based on measuring the weak values of system by selecting appropriate
pre- and post-selection states in order to generate polarization-path
quantum Cheshire cat states of two entangled photons. We found that
by properly choosing the pre- and post-select states of system, the
wave and particle attributes of two entangled photons can be spatially
separated, while keeping the validity of Bohr's complementarity principle.
We also proposed a feasible setup for the implementation of our thought
experiment with quantum optical platform.

As indicated in our result, we can control wave-particle attributes
of two entangled photons by adjusting the optical axis angle $\alpha$
of HWP. If $\alpha=0$ or $\alpha=\frac{\pi}{2}$, then we only have
pure wave attributes or pure particle attributes of two photons and
separate them spatially. Thus, it is very curious issue to do double-
slit experiment by using pure particle attributes of entangled photons
\citep{Hong98}. Furthermore, in the emerging field of counterfactual
communication \citep{PhysRevLett.110.170502,PhysRevA.92.052315},
information may be exchanged even without the transmission of quantum
systems. In this framework, entanglement also plays a vital role,
but the basic principles of such communication is still unclear. Thus,
it may interesting to use pure wave attributes of entangled photons
for understanding of the fundamental mechanism behind counterfactual
communication. At the same time, our results may also provide an alternative
method to enhance the precision of quantum metrological protocols.
In particular, in quantum metrology one would generally require that
the measurement of a physical quantity on an entangled state does
non disturb to other variable. Thus, the separation of entangled observables
by QCC and implement the related precision measurement may dramatically
increase its efficiency.

Another interesting point of our scheme is that it can scale up to
more multi-entangled photons cases. Since the parallel assembly of
$N$ single-photon WP toolboxes allows us to generate $N$-photon
wave-particle entangled states \citep{2017}, we may realize the spatial
separation of wave and particle attributes of $N$-photon wave-particle
entangled states by expand our schematics.

Finally, we anticipate that our theoretical proposal is feasible with
current technology and may be implemented in quantum optical labs
in the near future.
\begin{acknowledgments}
This work was supported by the National Natural Science Foundation
of China (No. 12365005)
\end{acknowledgments}

\bibliographystyle{apsrev4-1}
\bibliography{QCC-ref}

\end{document}